\documentclass{JASSS}
	
\usepackage[utf8]{inputenc}	
\title{The interplay between conformity and anticonformity and its polarizing effect on society.}

\reviewcopy{false} 

\author[1]{Patryk Siedlecki}
\author[1]{Janusz Szwabiński}
\affil[1]{University of Wrocław, Poland}

\author[2]{Tomasz Weron}
\affil[2]{Wrocław University of Technology, Poland}

\email{janusz.szwabinski@ift.uni.wroc.pl}

\usepackage{natbib}
	\setcitestyle{authoryear,round,aysep={}}
	
\begin{document}
\maketitle 

%%%%%%%%%%%%%%%%%%%%%%%%%%%%%%%%%%%%%%%%%%%%%%

% Abstract and keywords

\begin{abstract}
Simmering debates leading to polarization are observed in many domains. Although empirical findings show a strong correlation between this phenomenon and modularity of a social network, still little is known about the actual mechanisms driving communities to conflicting opinions. In this paper, we used an agent-based model to check if the polarization may be induced by a competition between two types of social response: conformity and anticonformity. The proposed model builds on the q-voter model~\citep{CAS:MUN:PAS:09} and uses a double-clique topology in order to capture segmentation of a community. Our results indicate that the interplay between intra-clique conformity and inter-clique anticonformity may indeed lead to a bi-polarized state of the entire system. We have found a dynamic phase transition controlled by the fraction $L$ of negative cross-links between cliques. In the regime of small values of $L$ the system is able to reach the total positive consensus. If the values of $L$ are large enough, anticonformity takes over and the system always ends up in a polarized stated.  Putting it the other way around, the segmentation of the network is not a sufficient condition for the polarization to appear. A suitable level of antagonistic interactions between segments is required to arrive at a polarized steady state within our model.
\end{abstract}


%%%%%%%%%%%%%%%%%%%%%%%%%%%%%%%%%%%%%%%%%%%%%%
% Start of  paragraph numbering. Please leave this untouched
\parano{}

\section{Introduction}
\label{sec:introduction}

In the last decades a lot of effort has been put into understanding and quantifying polarization within groups of people~\citep{ISE:86,SUN:02,ADA:GLA:05,DIX:WEI:07,LIV:SIM:ADA:ADA:11,GUE:MEI:CAR:KLE:13,MCC:DUN:11,MOU:SOB:01,DIM:EVA:BRY:96,MAO:06}. The reason is at least twofold. First,  polarization is relevant from the sociological point of view, because it often leads to segregation and conflits in the society~\citep{DIM:EVA:BRY:96}. Secondly, it plays an important role in opinion analysis and similar tasks~\citep{CAL:VEL:MEI:ALM:2011}. In particular, it may shed more light on polarized debates and help to predict their outcomes~\citep{WAL:91}.

Although the definition of polarization may differ slightly depending on the area of interest, the concept usually refers to a situation in which a group is divided into two opposing parties having contrasting positions~\citep{DIM:EVA:BRY:96}. For the sake of clarity this type of polarization is sometimes referred to as bi-polarization~\citep{MAS:FLA:13} to distinguish it from the group polarization phenomenon understood as the tendency for a group to make decisions that are more extreme than the initial inclination of its members~\citep{ISE:86,SUN:02}.

Simmering debates leading to polarization are observed in many domains. Topics such as global warming~\citep{MCC:DUN:11}, same-sex marriage and abortion~\citep{MOU:SOB:01} or the recent Syrian refugee crisis in Europe are known to drive people to extreme and opposing opinions. However, the world of politics is the leading domain where polarization is witnessed~\citep{GRU:ROY:14,ADA:GLA:05,MAO:06,WAU:PEI:FOW:MUC:POR:11}.

\begin{figure}
\centering
\includegraphics[scale=0.25]{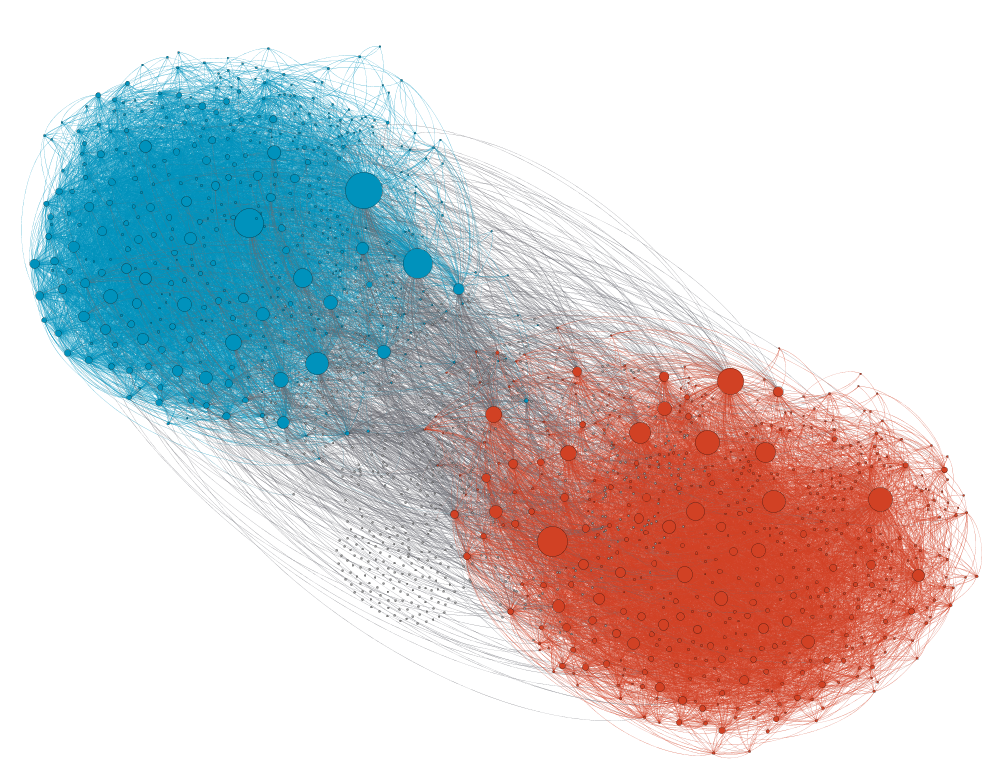}
\caption{(Color online) Linking paterns and discussion topics of political blogs over a period of two months preceding the U.S. Presidential Election of 2004. The colors reflect political orientation, red for conservative, and blue for liberal. The size of each blog reflects the number of other blogs that link to it. It turns out that more than 90\% of the links between blogs originating within either the conservative or the liberal community stay within that community. Data taken from~\citep{ADA:GLA:05}.}
\label{fig:us_blogs}
\end{figure}

With the raise of social media systems many "battles" on polemic issues have been moved to the Internet. As a side effect of this process, a vast amount of data on polarization related issues became (relatively) easy to access. Since then, this data is extensively studied by social scientists, complex networks experts, physicists and mathematicians with the goal to unveil some structural patterns that better capture the characteristics of polarization. An interesting example of this approach is the analysis of 2004 U.S. political blogosphere~\citep{ADA:GLA:05}. The authors studied linking patterns and discussion topics of political blogs over a period of two months preceding the U.S. Presidential Election of 2004. By making use of the network analysis methods they showed that there is the unmistakable division between the liberal and conservative political blogospheres (see Fig.~\ref{fig:us_blogs}). It turned out that more than 90\% of the links between blogs originating within either the conservative or the liberal community stay within that community. In other words, people turned to be starkly divided along partisan lines.

As can be seen in Fig.~\ref{fig:us_blogs}, the link network in the U.S. blogosphere example is highly modular. Further research on complex networks has shown that polarization may be indeed correlated to a network segmentation with high modularity~\citep{CON:RAT:FRA:GON:FLA:MEN:11,NEW:06,ZAC:77}. The presence of segmentation may be in turn related to the well-known social phenomenon of homophily, i.e. the tendency of individuals to associate and bond with similar others~\citep{MCP:SMI:COO:01,BRE:PRI:11}. Once the segments (cliques) are formed, their members are affected by different types of social influences. In this way, homogeneity within a group can be reached. At the same time, the groups may be pushed away towards extreme opinions, which results in a polarized state~\citep{SUN:02}. 

\begin{figure}
\centering
\includegraphics[scale=0.3]{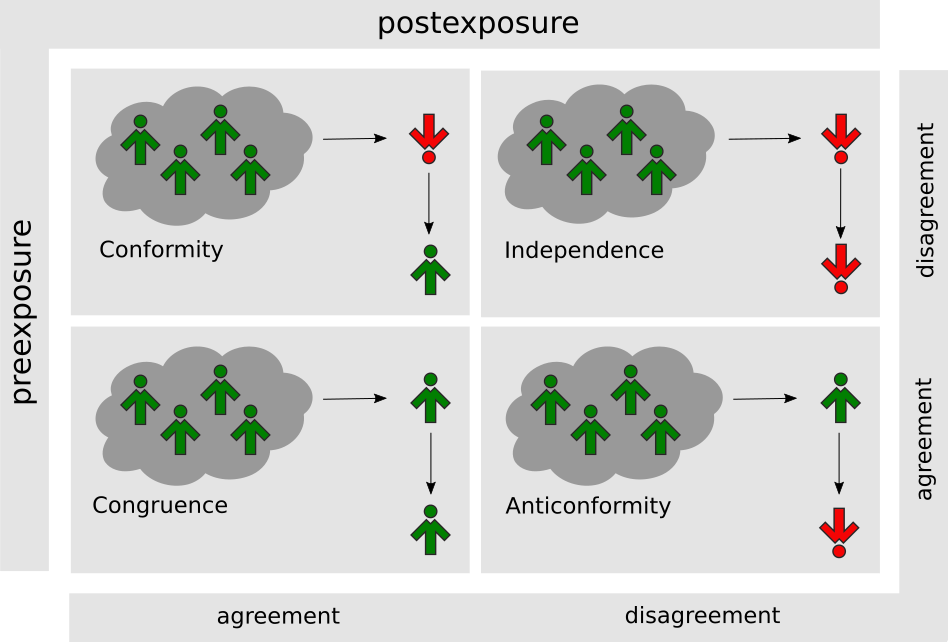}
\caption{Possible responses to the social influence according to the diamond model~\citep{WIL:63,NAI:86,NAI:DON:LEV:00,NAI:DOM:MAC:13}. Here presented in a q-voter model~\citep{CAS:MUN:PAS:09}. The source of influence is a group consisting of unanimous agents (schematically pictured as a cloud). The "up" and "down" spins (arrows) represent agents with two different opinions on a single issue.}
\label{fig:diamond}
\end{figure}

There have been several attempts to describe possible responses to social influence~\citep{ALL:65,ALL:34,SHE:35,CRU:62}. One of the most important achievements in this area is the diamond model first presented by Willis~\citep{WIL:63} and then formalized and extended by Nail and coworkers~\citep{NAI:86,NAI:DON:LEV:00,NAI:DOM:MAC:13}. The model is based on crossing two dichotomous variables: pre-exposure agreement or disagreement between a target and a source of influence and post-exposure agreement or disagreement between the target and the source~(Fig.~\ref{fig:diamond}).
Depending on the direction of target's movement, four possible responses to social influence may be identified within the model:
\begin{itemize}
\item \textbf{conformity} is the act of matching attitudes, beliefs and behaviours to group norms. Within the model shown in Fig.~\ref{fig:diamond} it is identified as the pre-exposure disagreement between a target and a source followed by the post-exposure agreement,
\item \textbf{independence} is the unwillingness to yield to the group pressure,
\item \textbf{anticonformity} refers to a situation when an individual consciously and deliberately challenges the position or actions of the group,
\item \textbf{congruence} may be understood as a type of conformity in which a target can "fit in" without having to change his/her views.
\end{itemize}

Although there are many different motivations to match or imitate others, and many different factors influence the level of conformity, this type of social response is ubiquitous in real societies~\citep{CIA:GOL:04,GRI:GOL:MOR:CIA:KEN:06}. Thus, it should be not surprising that in many attempts to simulate social systems by means of agent-based models conformity was the main driving force governing their dynamics~\citep{MAC:WIL:02,CAS:FOR:LOR:09}. 
In the voter model~\citep{CLI:SUD:73,HOL:LIG:75} for instance a target of the social influence takes the opinion of one of its neighbors. The model became popular for the natural interpretation of its rules in terms of opinion dynamics and has been intensively investigated in the recent years. The majority rule model~\citep{GAL:02} was proposed to describe public debates. Within this model in every iteration step a discussion group is selected at random and all its members take the majority opinion inside the group.  In the Sznajd model~\citep{SZN:SZN:00} a pair of neighboring agents sharing the same opinion impose it on their neighbors. The model turned out to be useful in different areas~\citep{CAS:FOR:LOR:09}. It was for example adopted  to describe voting behavior in elections.

The Sznajd model may be treated as a special case of the q-voter model~\citep{CAS:MUN:PAS:09}, which is one of the most general models of binary opinion dynamics. Within this model, introduced as a generalization of the voter model, $q$ randomly picked neighbors (with possible repetitions) influence a voter to change his opinion. The voter conforms to their opinion if all $q$ neighbors agree. If the neighbors are not unanimous, the voter can still flip with a probability $\epsilon$. One of the strengths of the model is that it can be justified from the social point of view. The unanimity rule 
is in line with a number of social experiments. For instance, it has been observed that a larger group with a nonunanimous majority is usually less efficient in terms of social influence than a small unanimous group~\citep{MYE:13}. Moreover, Asch found that conformity is reduced dramatically by the presence of a social supporter: targets of influence having a partner sharing the same opinion  were far more independent when opposed to a seven person majority than targets without a partner opposed to a three person majority~\citep{ASC:55}.

All the models presented above, and many others using conformity as a main type of social influence (see~\citep{CAS:FOR:LOR:09} for further reference) share a common feature - complete consensus is their steady state. However, in real societies a complete unanimity is rather hardly reached~\citep{HUC:07}. To make the models more realistic, in the recent years several attempts have been undertaken to introduce other responses to social influence into the opinion dynamics. For instance, Galam introduced non-social states~\citep{GAL:MOS:91,GAL:97}, inflexible agents~\citep{GAL:JAC:07} and contrarians~\citep{GAL:04} into his models. While the first two modifications are different manifestations of independence identified within the diamond model (Fig.~\ref{fig:diamond}), the latter one fits into the anticonformity category. Indeed, non-social agents and inflexibles either change their opinions while ignoring their environment or they keep their opinions always unchanged. The contrarians on the other hand are able to adopt an opinion opposite to the prevailing choice of the others. Sznajd-Weron and coworkers considered in a number of papers the interplay between conformity and other types  of social influence in both the original Sznajd model~\citep{SZN:TAB:TIM:11} and the q-voter model~\citep{NYC:SZN:CIS:12,NYC:SZN:13,PRZ:SZN:WER:14,SZN:SZW:WER:WER:14,SZN:SZW:WER:14,JED:SZN:SZW:16}. It turned out that accounting for responses other than conformity significantly changes the output of the models. Not only the time evolution changes, but also the steady states are different and phase transitions of different types appear in the systems.

Although agent-based models have become one of the most powerful tools available for theorizing about opinion dynamics in general~\citep{LEI:14}, there are only few attempts to apply this kind of models to the phenomenon of polarization. 
For instance, French~\citep{FRE:56}, Harary \citep{HAR:59} and Abelson~\citep{ABE:64} showed that consensus must arise in populations whose members are unilaterally connected unless the underlying social network is separated. According to Axelrod~\citep{AXE:97} local convergence may lead to global polarization. A number of papers has been devoted to explain polarization within the social balance theory, i.e. by accounting for sentiment in dyadic and/or triadic relations in social networks~\citep{MAR:KLE:KLE:STR:11,TRA:DOO:LEE:13}. In other computer experiments it has been shown that bridges between clusters in a social network (long-range ties) may foster cultural polarization if homophily and assimilation at the micro level are combined with some negative influence, e.g. xenophobia and differentiation~\citep{MAC:KIT:FLA:BEN:03, SAL:06}. 
Finally, from other studies it follows that polarization may be also induced by mass media communication~\citep{MCK:SHE:06}. Nevertheless, our knowledge of the mechanisms governing the dynamics of polarization and in particular the role of social influence  remains sketchy. Further effort is needed to better understand how separation occurs in social groups. Thus, our goal in this paper is to show within an agent-based model that a combination of social responses identified in the diamond model (Fig.~\ref{fig:diamond}) can lead to social polarization in a modular society. To this end we will use a modified $q$-voter model~\citep{CAS:MUN:PAS:09} as our modelling framework. The reason is at least two-fold. It uses conformity as the main driving force, which makes sense from the social psychology point of view, because the tendency to match or imitate others  is omnipresent in social systems~\citep{CIA:GOL:04,GRI:GOL:MOR:CIA:KEN:06}. Moreover, the model is general enough to easily account for other types of social responses.

From the social balance theories it follows that both positive and negative ties are needed for the segregation to emerge and to prevail~\citep{TRA:DOO:LEE:13}. The positive ties may be related to the already discussed conformity. In terms of the diamond model (Fig.~\ref{fig:diamond}) negative ties are in turn best described by anticonformity. Thus, we will add this type of response to our model to check how the interplay between conformity and anticonformity impacts its dynamics.  One should however have in mind that the assumption on negative influence is still a subject of an intense debate. There are several models able to explain polarization without any kind of negative influence, for example the argument-communication theory of bi-polarization~\citep{MAS:FLA:13} or the bounded-confidence model~\citep{HEN:KRA:02}. Some empirical studies on negative influence do not provide convincing support for this assumption as well~\citep{KRI:BAR:07}.

The $q$-voter model we have chosen as our modelling framework belongs to the class of binary opinion models, i.e. models with agents characterized by a single dichotomous variable, $S_i =\pm 1$ ($i=1,\dots,N$). One might find this approach very unrealistic at first glance, because the opinions of individuals on specific subjects are expected to vary gradually and be described by a continuous variable~\citep{FRE:56,HAR:59,HEN:KRA:02,MAS:FLA:13}. From empirical data it follows that often this is not the case. It turns out for instance that the distribution of opinions on important issues measured on some multivalued scale is typically bimodal and peaked at extreme values~\citep{LEW:NOW:LAT:92,STO:GUT:SUC:LAZ:STA:CLA:50}. In other words, in some situations the most important characteristics of the system under investigation may be  already captured by the relative simple models of binary opinions.

The paper is organized as follows. In the next section we introduce our model and shortly describe how it differs from the original $q$-voter model. Then, we will investigate the model by means of agent-based simulations. Finally some conclusions will be presented.

\section{Model}
\label{sec:model}

Many data on social networks are characterized by a semantic unicity, meaning that opinions and interactions of networks' members are restricted to a single domain or topic~\citep{GUE:MEI:CAR:KLE:13}. Moreover, very often those opinions may be interpreted as simple "yes"/"no", "in favour of"/"against" or "adopted"/"not adopted" answers~\citep{WAT:DOD:07}. Thus, we decided to focus our attention on binary opinion models with a single trait.

We consider a set of $2N$ agents, each of whom has an opinion on some issue that at any given time can take one of two values: $S_i=-1$  or $S_i=1$  for $i=1,2,\dots,2N$. Following~\citep{PRZ:SZN:WER:14} and~\citep{NYC:SZN:13} we will call these agents spinsons to reflect their dichotomous nature originating in spin models of statistical physics and humanly features and interpretation (\emph{spinson} = \emph{spin}+per\emph{son}).

\begin{figure}
\centering
\includegraphics[scale=0.4]{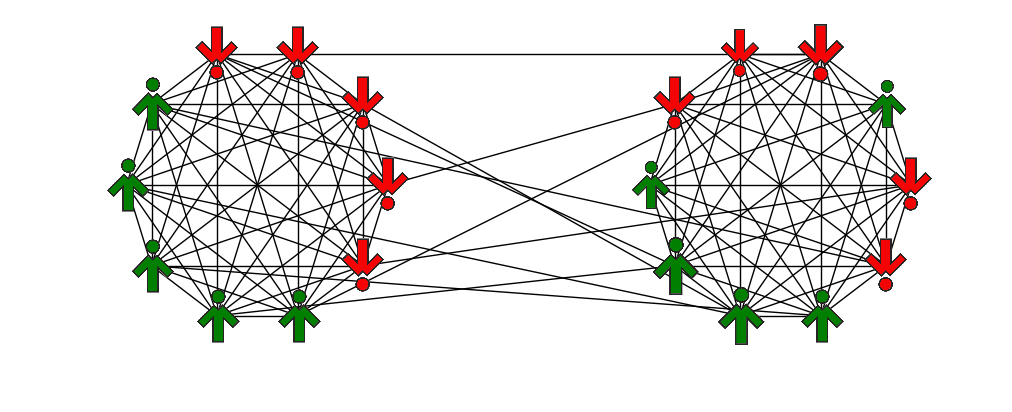}
\caption{An example of a double-clique network. The network consists of two separate complete graphs (cliques) with some cross links between them. The parameters of this particular model: $N=10$ nodes in each clique and $L=0.1$ (i.e. $L\times N^2=10$ cross links).}
\label{fig:two_clique}
\end{figure}

From social networks analysis it follows that the existence of two segregated groups may foster polarization~\citep{CON:RAT:FRA:GON:FLA:MEN:11,NEW:06,ZAC:77}. Since our goal is to investigate the role of social influence and long-range ties rather than the mechanism leading to network modularity, we will assume that our social network is already modular. Thus we put our agents on a so called double-clique topology~\citep{SOO:ANT:RED:08}. An example of such a network is shown in Fig.~\ref{fig:two_clique}. It consists of two separate complete graphs of $N$ nodes, connected with $L\times N^2$ cross links. Here, $N^2$ is the maximal number of cross links between the cliques (when every agent from one clique of size $N$ is connected to all agents from the other clique of the same size). $L$ is the fraction of existing cross links. It is one of the parameters of our model describing the extent of inter-clique connectivity.

Conformity and anticonformity were identified as opposites in the diamond model~\citep{NAI:86,NAI:DON:LEV:00,NAI:DOM:MAC:13} shown in Fig.~\ref{fig:diamond}. However, while conformity is produced by a need of social acceptance and anticonformity is motivated rather by negativism and hostility, these responses may be treated as mirror images of each other. Both reflect dependence on the group, which is a positive
reference group for conformity and a negative one for anticonformity~\citep{KRE:CRU:BAL:62}. Moreover, in many settings multiple sources of norms are possible. As a consequence, labelling of the responses is relative, because conformity to one source can at the same time be anticonformity to another. For instance, a teenager's conformity to peers is very often anticonformity to his parents~\citep{NAI:DON:LEV:00}. To elaborate on that issue we will assume within our model, that a spinson strives for agreement within his own clique and simultaneously anticonforms to individuals from the other clique.

\begin{figure}
\centering
\includegraphics[scale=0.4]{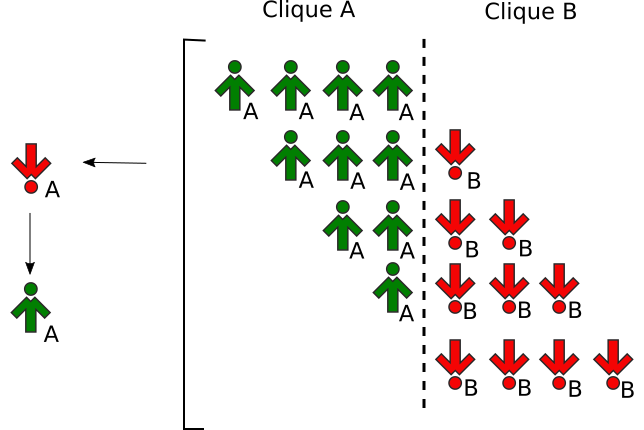}%{influence}
\caption{Possible choices of the influence group in case $q=4$, that lead to an opinion flip of a spinson in clique $A$ being initially in state $S=-1$. The target receives ``signals'' emitted by the members of the influence group. A signal is just a state of a member when coming from target's clique or its inverted state otherwise. The target changes its opinion only if all members of the influence group emit the same signal.}
\label{fig:influence}
\end{figure}

We use Monte Carlo simulation techniques with a random sequential updating scheme to investigate the model (see Appendix A). Each Monte Carlo step in our simulations consists of $2\times N$ elementary events, each of which may be divided into the following substeps:
\begin{enumerate}
\item Pick a target spinson at random.
\item Build its influence group by randomly choosing $q$ neighboring agents.
\item Convert the states of the neighbors into signals that may be received by the target. Assume that the signals of the neighbors from target's clique are equal to their states. Invert the states when from the other clique.
\item Calculate the total signal of the influence group by summing up individual signals of its members.
\item If the total signal is equal to $\pm q$ (i.e. all group members emit the same signal), the target changes its opinion accordingly (see Fig.~\ref{fig:influence}). Otherwise nothing happens.
\end{enumerate}
Note that our model is nothing but a modification of the $q$-voter with $\epsilon=0$ and an additional social response of spinsons. Indeed, in the original $q$-voter model~\citep{CAS:MUN:PAS:09} conformity is the only driving force in the system: for a randomly chosen target  a group of $q$ neighbors is picked at random as well. If all neighbors are in the same state, the target adopts their opinion. Otherwise the target may flip its state with an independent probability $\epsilon$. 

Since the target in our model may act as both conformist and anticonformist at the same time, the concept of unanimity of the influence group from the original $q$-voter model requires some modifications. To this end we have introduced the notion of ``signals`` in the above steps. A signal is just a state of a neighbor when coming from target's clique or its inverted state otherwise. The target changes its opinion only if all  members of the influence group emit the same signal. This allows us to treat both conformity and anticonformity on the same footing within the $q$-voter framework. Working with signals has a consequence - different configurations of the influence group may result in the same signal, as shown  in Fig.~\ref{fig:influence}.

\section{Results}
\label{sec:res}

All results presented in this section were obtained  with Monte Carlo simulations on the double clique topology shown in Fig.~\ref{fig:two_clique}.  All simulations were run until equilibrium was reached. If not explicitly stated otherwise, results for each set of the control parameters were averaged over 10000 independent trials in order to get reasonable statistics. 

As already indicated in the previous section, our model has three control parameters: the size of a clique $N$, the fraction of existing cross-links $L$ and the size of an influence group $q$. These parameters were systematically varied to generate different initial model configurations. We proceeded as follows: once the values of $N$ and $q$ were set, we varied $L$ from $0$ to $1$ with step $0.01$ in order to get some impression on the impact of antagonistic cross-links on the dynamics. Then we changed the values of $N$ and/or $q$ and repeated the procedure. In this way the role of the system and influence group size could be analysed as well. The parameters and their values are summarized in Table~\ref{tab:summary}.

\begin{table}
\centering
\begin{tabular}{l|l|l}
\hline\hline
Parameter & Description & Values\\
\hline\hline
$N$ & Number of agents in one clique & $100,~200,~400,~500,~1000,$\\
& & $2000,~4000$\\
$L$ & Fraction of existing cross links & from $0$ to $1$ with step $0.01$\\
$q$ & Size of the influence group & $2,~3,~4,~5,~6$\\
\hline\hline
\end{tabular}
\caption{Summary of the model parameters.}
\label{tab:summary}
\end{table}

\subsection{Average opinion}

For the models of opinion dynamics, a good measure of the macroscopic state of a given system is an average opinion $m$, defined in analogy to magnetization in spin systems as
\begin{equation}
m = \frac{1}{N}\sum_i S_i,
\end{equation} 
where $S_i$ is the opinion of the $i$-th agent and $N$ is the number of agents. Assuming $S_i=1$ as an "in favor of" opinion on a given issue, the interpretation of the average opinion is straightforward:
\begin{itemize}
\item $m=1$ - positive consensus, i.e. all spinsons are "in favor of" that issue,
\item $0<m<1$ - partial positive ordering, i.e. the majority of the spinsons is "in favor of",
\item $m=0$ - no ordering in the system, the "in favor of" and "against" groups are equally populated,
\item $-1<m<0$ - partial negative ordering, i.e. the majority of the spinsons is "against" the issue,
\item $m=-1$ - negative consensus, all agents are "against" the issue.
\end{itemize} 

In case of the double-clique topology, it may be more insightful to look at the average opinion of a single clique rather than that of the entire system. Thus, we will slightly modify the above definition:
\begin{equation}
m_X = \frac{1}{N_X}\sum_i S_{X,i},~~~~X=A,B
\end{equation} 

\subsection{Note on initial conditions}

\begin{figure}
\centering
\includegraphics[scale=0.25]{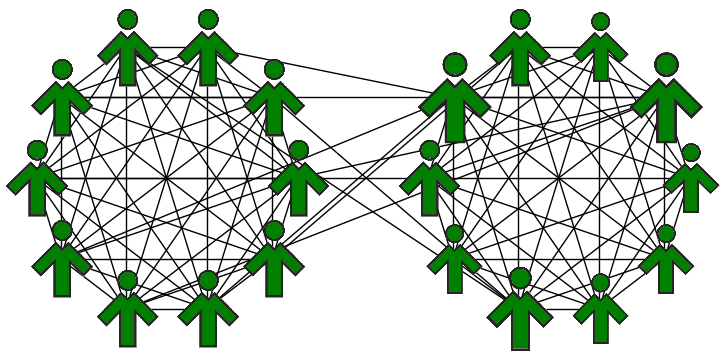}  \ \hfill%{init_1.png}
\includegraphics[scale=0.25]{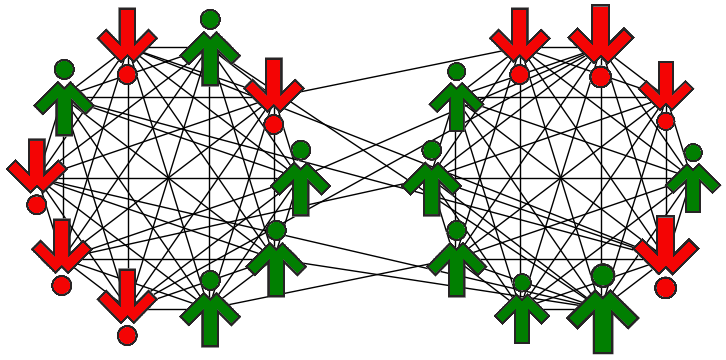}%{init_0.png}
\caption{Two different initial configurations: total positive consensus  (left plot) and no ordering in the system (right). They will be referred to as Scenario I and Scenario II, respectively.}
\label{fig:init}
\end{figure}

In our simulations we looked at two different initial conditions: (1) total positive consensus, i.e. $m_{A,B} (0) = 1$ (left panel in Fig.~\ref{fig:init}) and (2) no ordering in the system, $m_{A,B} (0) = 0$ (right panel).

These configurations may be treated as results of two different scenarios. In the first one we assume that two cliques with a natural tendency to disagree with each other evolve at first independently.  They get in touch by chance and establish some cross-links to the other group once they both reached consensus on a given issue.

In the second scenario the community is already divided into two loosely connected antagonistic groups. When a new issue appears, the agents form their views based on their tastes before the group pressure sets in. As a result, both "in favor of" and "against" opinions are initially rather randomly distributed over the system resulting in no ordering in the cliques. 

These two different situations will be hereafter referred to as Scenario I and Scenario II.

\subsection{Time evolution of average opinion}
\label{subsec: evolution}

\begin{figure}
\centering
\includegraphics[scale=0.4]{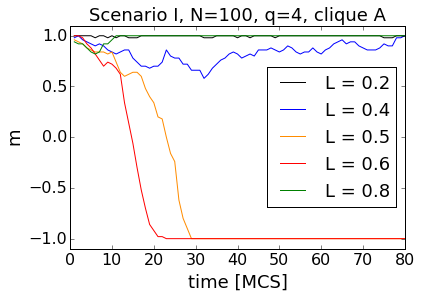}%{time_evol_single_11_A.png}
\includegraphics[scale=0.4]{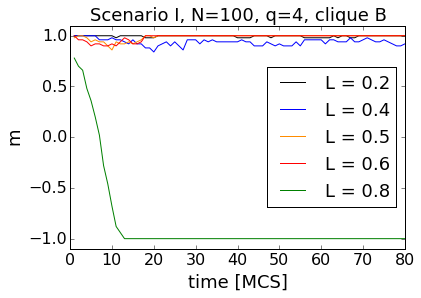}%{time_evol_single_11_B.png}
\caption{(Color online) Time evolution of the average opinion in clique $A$ (left plot) and clique $B$ (right plot) for different values of $L$. The parameters of the model: $N=100$ nodes in each clique and $q=4$ agents in a group of influence. The results were obtained from single runs starting from the total positive consensus (Scenario I).}
\label{fig:time_evol_single}
\end{figure}

Let us first analyse the time evolution of the average opinion starting from the total positive consensus (Scenario I). Trajectories for different values of $L$ obtained from single runs are shown in Fig.~\ref{fig:time_evol_single}.
In this particular example each clique consisted of $N=100$ agents and the size of the influence group was set to $q=4$. First of all, we see that each clique may end up in one of two possible asymptotic states: positive or negative consensus. Moreover, the average opinions in the cliques may have either equal or  opposite signs. Note that the asymptotic state is reached rather quickly. From the example trajectories shown in Fig.~\ref{fig:time_evol_single} only the one for $L=0.4$  has not arrived at its asymptotic value after 80 Monte Carlo steps and requires longer times to reach its equilibrium. This issue will be addressed again in par.~\ref{par: times}.\label{par: evo}

\begin{figure}
\centering
\includegraphics[scale=0.4]{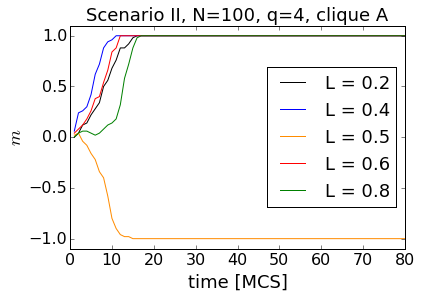}%{time_evol_single_00_A.png}
\includegraphics[scale=0.4]{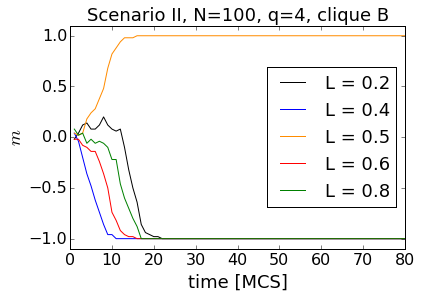}%{time_evol_single_00_B.png}
\caption{(Color online) Time evolution of the average opinion in case of no initial ordering (Scenario II). See caption of Fig.~\ref{fig:time_evol_single} for further details.}
\label{fig:time_evol_single_00}
\end{figure}

The case of no initial ordering (Scenario II) is similar in the sense, that the two asymptotic states are the same and both cliques reach them quickly. However, actual trajectories differ from the previous case, as shown in Fig.~\ref{fig:time_evol_single_00}.

\begin{figure}
\centering
\includegraphics[scale=0.4]{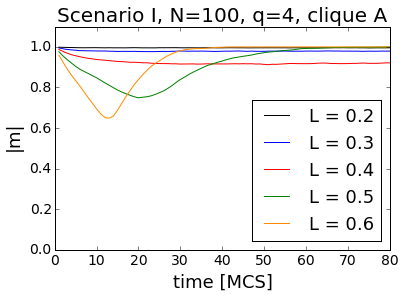}%{time_evol_aver_11_A.png}
\includegraphics[scale=0.4]{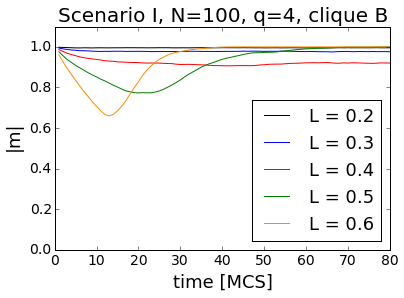}%{time_evol_aver_11_B.png}
\caption{(Color online) Time evolution of the average opinion (its absolute value) in clique A and B averaged over 10000 independent runs for different values of $L$. The parameters of the model: $N=100$ nodes in each clique and $q=4$ agents in a group of influence. Initial conditions from Scenario I.}
\label{fig:time_evol_aver_11}
\end{figure}

\begin{figure}
\centering
\includegraphics[scale=0.4]{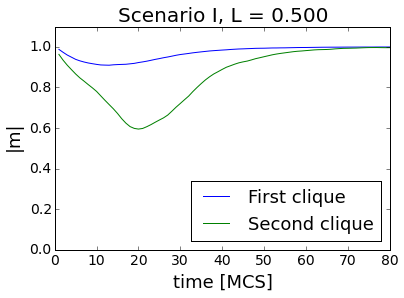}%{time_evol_aver_11_fs.png}
\includegraphics[scale=0.4]{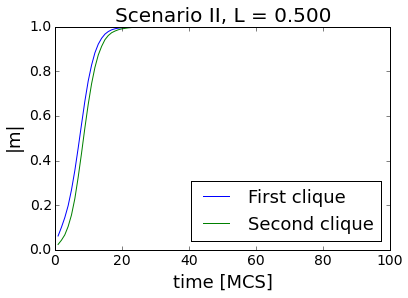}%{time_evol_aver_00_fs.png}
\caption{(Color online) Comparison between averaged time trajectories of cliques arriving at the asymptotic state as the first and the second ones. The parameters of the model: $N=100$ nodes in each clique and $q=4$ agents in a group of influence.}
\label{fig:time_evol_aver_fs}
\end{figure}

Due to the stochastic nature of the simulations, there is usual a fair amount of run-to-run variance in our computer experiments. Thus, we cannot expect to conclude everything from single runs and should also look at the averages over independent runs. In Fig.~\ref{fig:time_evol_aver_11}, time trajectories averaged over 10000 runs are shown for both cliques in Scenario I.
It seems that on average both cliques evolve in the same way (true for Scenario II as well). This result indicates that the probability of reaching the asymptotic state as first is equal for both cliques. Actually, this is something one could expect, because our model is initially invariant against interchanging of the clique labels. Differences between the cliques are induced dynamically due to the stochastic nature of the simulation. Thus,  we decided to do the averaging of the results in a slightly different way: in every run we marked the clique reaching as first its asymptotic state and took the averages over runs separately for those first cliques and for the other ones.
In doing so we can indeed observe some differences in the time evolution toward the asymptotic states, as shown in Fig~\ref{fig:time_evol_aver_fs}. 

As you may already noted, in Figs.~\ref{fig:time_evol_aver_11}-\ref{fig:time_evol_aver_fs} we looked at the absolute value of the average opinion rather than the opinion itself. The reason for that is the already mentioned run-to-run variance due to the stochastic dynamics. In Scenario II for instance, for a given set of parameter values a single run could reach both asymtotic states with equal probability. Thus averaging the opinion over independent runs would therefore yield 0 in all cases. In Scenario I the situation was similar, provided the parameter $L$ was large enough.  

\subsection{Correlations between cliques}

The final remark in the previous section indicates that the competition between conformity and anticonformity may lead to a phase  transition, at least within Scenario I. To elaborate on that issue we will measure the correlation between final states of the cliques, defined in the following way:
\begin{equation}
\langle m_A^\infty m_B^\infty\rangle = \frac{1}{N_{runs}}\sum_i m_{A,i}^\infty m_{B,i}^\infty
\label{eq: corr}
\end{equation} 
Here, $m_{A,i}^\infty$ and $m_{B,i}^\infty$ denote the final states of clique A and B in the $i$-th run for the same set of parameters $(N,q,L)$, $N_{runs}$ is the number of runs we average over and the summation index goes over all runs.  For the sake of convenience we will neglect the superscript $\infty$ in the plots presented below.

\begin{figure}
\centering
\includegraphics[scale=0.4]{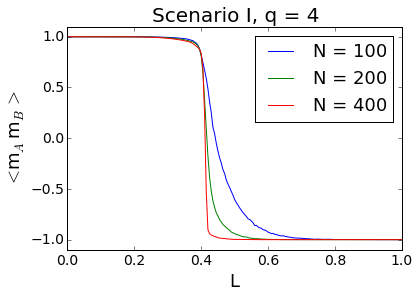}%{corr_n_11.png}
\includegraphics[scale=0.4]{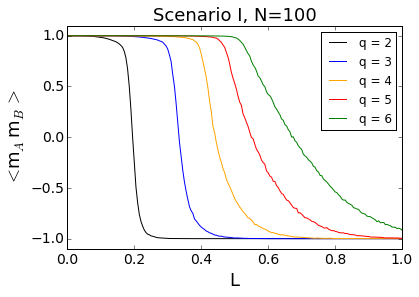}%{corr_q_11.png}
\caption{(Color online) Correlations between cliques as functions of $L$ for different values of $N$ (left) and $q$ (right) within Scenario I.}
\label{fig:corr11}
\end{figure}

Correlations between cliques as functions of $L$ for different values of $N$ (left plot) and $q$ (right plot) within Scenario I are shown in Fig.~\ref{fig:corr11}.
Indeed, there appears a phase transition in the system. For values of $L$ smaller than a critical value ($\approx 0.4$ for $q=4$) both cliques always end up in the total positive consensus. In other words, in this regime the intra-clique conformity wins with the inter-clique anticonformity and both communities are able to maintain their initial positive consensus. If the value of $L$ is larger than the critical one, the anticonformity induced effects take over and the whole system ends up in a polarized state.

As we can see from Fig.~\ref{fig:corr11}, the transition depends only little on the system size. The bigger $N$, the sharper is the transition, but the critical point remains roughly the same. Moreover, for clique sizes bigger than $N=400$ no changes in $\langle m_A^\infty m_B^\infty\rangle$ have been observed.

As far as the impact of the influence group is concerned, we see that the critical point shifts with increasing $q$ towards higher values of $L$. Thus, the smaller the influence group, the less cross-links are needed to polarize the society.

\begin{figure}
\centering
\includegraphics[scale=0.4]{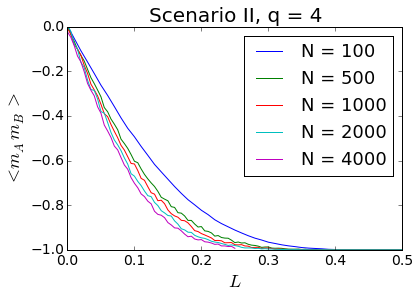}%{corr_n_00.png}
\includegraphics[scale=0.4]{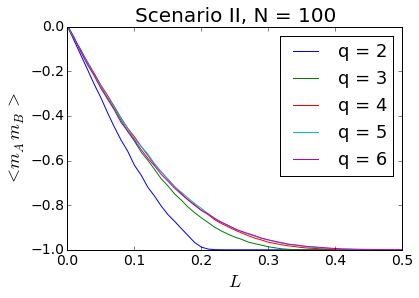}%{corr_q_00.png}
\caption{(Color online) Correlations between cliques as functions of $L$ for different values of $N$ (left) and $q$ (right) within Scenario II.}
\label{fig:corr00}
\end{figure}

In Fig.~\ref{fig:corr00} correlations between cliques are shown for Scenario II.
In this case we also observe a kind of phase transition. Below a critical value of $L$ the correlation function is negative and greater than $-1$. According to Eq.~(\ref{eq: corr}) it means that the system ends up in the polarized state in most of the cases, but some of the independent runs lead to consensus. Above the critical value the cliques are always polarized.
Compared to the previous case the transition is not very sharp. The critical value is smaller and it depends little on both the clique size $N$ and the influence group size $q$.

\subsection{Relaxation times}
\label{subsec: times}

As already mentioned in Sec.~\ref{subsec: evolution}, the cliques usually reach their asymptotic states very quickly. To investigate that issue in more detail, we look now at the relaxation times $\langle\tau\rangle$, i.e. average times needed to reach consensus within each clique.

\begin{figure}
\centering
\includegraphics[scale=0.4]{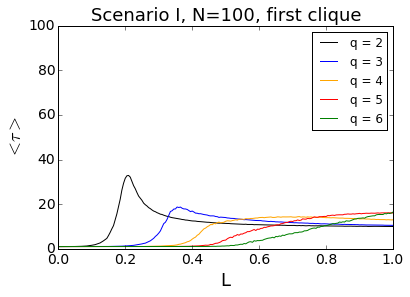}%{relaxation_time_100_first_11.png}
\includegraphics[scale=0.4]{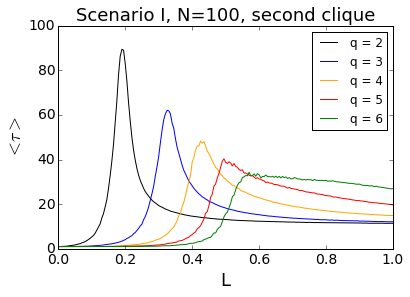}%{relaxation_time_100_second_11.png}
\caption{(Color online) Relaxation times within Scenario I as functions of $L$. 
Results were averaged separately for cliques reaching their asymptotic states as first (left plot) and as second (right plot). }
\label{fig: reltime11}
\end{figure}

In Fig.~\ref{fig: reltime11} relaxation times for different values of $q$ and $N=100$ within Scenario I are shown.  
The results confirm our impression from par.~\ref{par: evo} - the relaxation times are indeed rather short and  decrease with $q$. Moreover, critical slowing down may be clearly seen in the plots. The times peak  at the critical values of $L$ meaning that near the critical point it takes longer for the system to arrive at its final state. This phenomenon is know from statistical physics. It occurs because system's internal stabilizing forces become weaker near the transition point, at which they suddenly propel the system toward a different state. In our model, conformity is the stabilizing force at small values of $L$, and anticonformity at the large ones. Near the critical point the impact of both forces on the system becomes comparable, the system experiences abrupt changes in its state and it just takes longer till one of the forces takes over and the system 'decides' to evolve toward a final state.\label{par: times}

\begin{figure}
\centering
\includegraphics[scale=0.4]{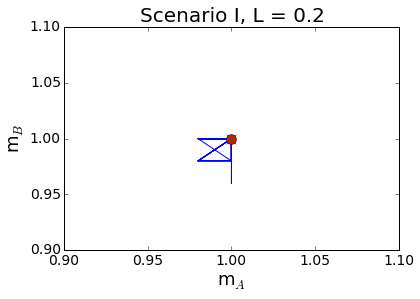}%{phase_space_02_11.png}
\includegraphics[scale=0.4]{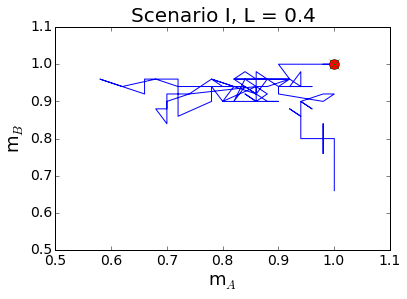}\\%{phase_space_04_11.png}
\includegraphics[scale=0.4]{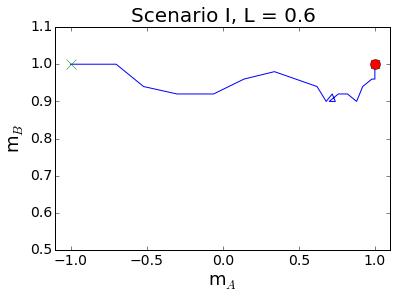}%{phase_space_06_11.png}
\caption{(Color online) Trajectories in phase space for Scenario I and different values of $L$. Red dot indicates the initial state, green cross the final one. Note the differences in the scales across the plots. Parameters of the model: $N=100$ and $q=4$. }
\label{fig:phasespace11}
\end{figure}

To illustrate that phenomenon three different trajectories of the system in the phase space are shown in Fig.~\ref{fig:phasespace11}.
For $L=0.2$ the system is in the regime, in which the internal conformity plays the most important role. As a result, both cliques stay most of the time near their initial values and reach the final state quickly. Similarly, for $L=0.6$ the anticonformity plays the crucial role for the opinion dynamics  and the system is driven straight towards polarization. In the $L=0.4$ case the value of $L$ is close to the critical value. Both forces have similar impact and the system performs a kind of a random walk in the phase space before it hits the asymptotic state.
 
\begin{figure}
\centering
\includegraphics[scale=0.4]{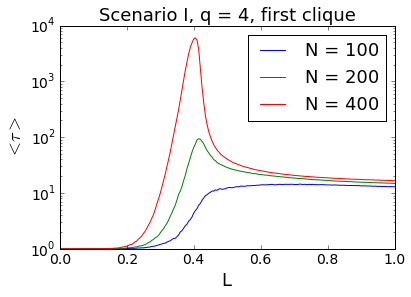}%{relaxation_time_n_first_11.png}
\includegraphics[scale=0.4]{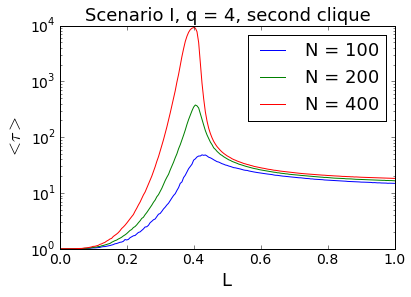}%{relaxation_time_n_second_11.png}
\caption{(Color online) Relaxation times of the first and second clique in Scenario I for different values of $N$. See the caption of Fig.~\ref{fig: reltime11} for the explanation of the clique labels. Note the logarithmic scale on the vertical axis.}
\label{fig:reltimeN11}
\end{figure}

To conclude the discussion on Scenario I, let us look at the dependence of the relaxation times on the size of the cliques. Corresponding results are shown in Fig.~\ref{fig:reltimeN11}.
We see that the clique size plays an important role in the vicinity of the critical point. The peak value of the relaxation time increases significantly with $N$, which has a negative impact on the computing times during simulations.

\begin{figure}
\centering
\includegraphics[scale=0.4]{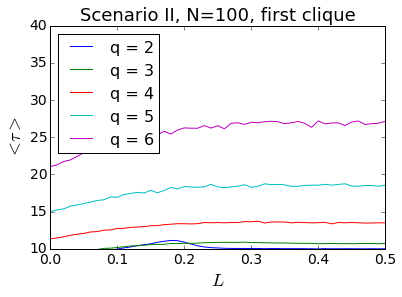}%{relaxation_time_100_first_00.png}
\includegraphics[scale=0.4]{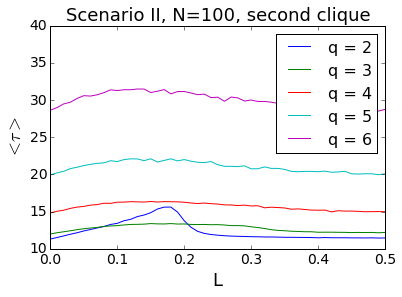}%{relaxation_time_100_second_00.png}
\caption{(Color online) Relaxation times within Scenario II as functions of $L$. See the caption of Fig.~\ref{fig: reltime11} for the explanation of the clique labels. }
\label{fig:reltime00}
\end{figure}

As for the Scenario II, relaxation times for different values of $q$ are shown in Fig.~\ref{fig:reltime00}. 
Compared to Scenario I, the relaxation times are now shorter and the differences between the first and the second clique are much smaller. A well pronounced peak is visible only for $q=2$. For bigger sizes of the influence group the maxima are smeared out or even unnoticeable.  

\begin{figure}
\centering
\includegraphics[scale=0.4]{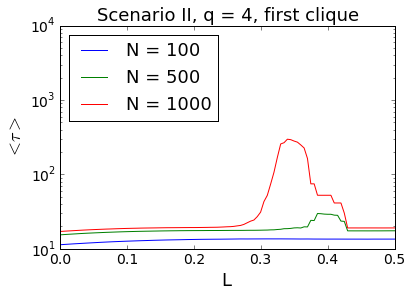}%{relaxation_time_n_first_00.png}
\includegraphics[scale=0.4]{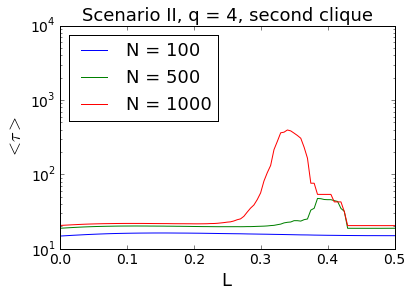}%{relaxation_time_n_second_00.png}
\caption{(Color online) Relaxation times of the first and second clique in Scenario II for different values of $N$. See the caption of Fig.~\ref{fig: reltime11} for the explanation of the clique labels. Note the logarithmic scale on the vertical axis.}
\label{fig:reltimeN00}
\end{figure}

The dependence of the relaxation times on the size of the cliques is similar to the results for Scenario I (see Fig.~\ref{fig:reltimeN00}). Although still shorter, the maximal relaxation times increase with $N$.
Moreover, for $N$ larger than 100  well pronounced peaks appear in the curves, indicating again a slowing down near the critical point. 

\section{Conclusions}
\label{sec:con}

Available data about different debates on polemic issues indicate that bi-polarization is a pattern often observed in communities~\citep{GUE:MEI:CAR:KLE:13}. Moreover, various studies have shown that this polarization is correlated to the segmentation of the social network describing interactions in these communities~\citep{NEW:06}. Having these findings  in mind, we have proposed a variant of the q-voter model~\citep{CAS:MUN:PAS:09} with $\epsilon=0$ and a second type of social response of spinsons. In addition to conformity as the main force driving agents to change their opinions in accordance to the group pressure we have introduced negative social ties characterized by anticonformity, as proposed by Nail and coworkers~\citep{NAI:86,NAI:DON:LEV:00,NAI:DOM:MAC:13}. Since we were primarily interested in the interplay between different types of social response, we have assumed that the underlying social network is already segmented and have modelled it by the double-clique topology (see Fig.~\ref{fig:two_clique}).   

Our results indicate that the competition between conformity and anticonformity may indeed lead to a polarized state of the entire system. We have found a dynamic phase transition controlled by the fraction $L$ of cross-links between cliques. In the regime of small values of $L$ (i.e. only few antagonistic cross-links) conformity within a clique seems to be the main driving force and the system is able to reach the total consensus always in Scenario I and sometimes in Scenario II. If the values of $L$ are large enough, anticonformity takes over and the system always ends up in a polarized stated. Near the critical point the impact of these two forces is comparable resulting in a significant slowing down of the dynamics. Putting it the other way around, the segmentation of the network is not a sufficient condition for the polarization to appear. A suitable level of antagonistic interactions between segments is required to arrive at a bi-polarized steady state. Moreover, the critical value of $L$ increases with $q$. For bigger influence groups  the social pressure induced by intra-clique conformity is stronger and more inter-clique anticonformity is needed to compensate it. Thus, polarization is harder to achieve in communities, the members of which try to adjust their opinions to large groups of contacts. 

There are several potentially interesting extensions to the present model. For instance, it is already known that including noise to models of opinion dynamics may change their predictions~\citep{KLE:EGU:TOR:MIG:03}. Thus, it could be very informative to check how robust  our model is to the introduction of noise. Adding it to our modeling framework, e.g. in form of independence of agents~\citep{SZN:SZW:WER:14}, is very easy. We expect that at low noise rates the results should be substantially the same, whereas at higher rates the point $(m_A=0,m_B=0)$ should become a stable fixed point. However, as pointed out by Axelrod, intuition is not a very good guide for predicting what even a very simple opinion dynamics model will produce~\citep{AXE:97}. Hence, it is worth to check the accuracy of our expectations in a future work.

In the present studies we worked with the assumption that the network segmentation is (at least to some extent) responsible for the polarization. However, one cannot actually rule out the opposite possibility that the segmentation is induced or intensified by polarized opinions. The casual connection between the network segmentation and the polarization is therefore another interesting aspect worth to address in future studies.

\section{Acknowledgements}

This work was partially supported by funds from the Polish National Science Centre (NCN) through
grant no. 2013/11/B/HS4/01061.

%%%%%%%%%%%%%%%%%%%%%%%%%%%%%%%%%%%%%%%%%%%%%%
% End of  paragraph numbering. Please leave this untouched
\endparano

\section{Appendix A: Pseudocode of the simulation}
\label{code}

\begin{verbatim}
for every independent run:

    generate two complete graphs (cliques) of size N
    generate L*N**2 random connections between cliques 
    
    initialize states of agents (nodes):
        if Scenario I:
            initialize all states in both cliques to "1"
        if Scenario II:
            for each clique:
                choose an agent randomly
                draw a random number r from a uniform distribution
                if r smaller than 0.5:
                    initialize agent's opinion to "1"
                else:
                    initialize agent's opinion to "-1"
        


    for every time step:
        try as many times as the total number of agents (i.e. 2*N):
            choose an agent (target) randomly:
                choose q of its neighbors randomly
                determine signals of the neighbors:
                    if neighbors from target's clique:
                        take their states as signals 
                    else:
                        invert their states
                           
                calculate total signal of the influence group
                if total signal equal to the +/- size of the group:
                    adjust target's state (if necessary)                  
        
        calculate average opinion for each clique

average results over runs 
\end{verbatim}

%%%%%%%%%%%%%%%%%%%%%%%%%%%%%%%%%%%%%%%%%%%%%%

% REFERENCES.
% The JASSS bibliographic style file (jasss.bst) is included in the bundle. Please use BibTeX, not BibLaTeX.
% Use natbib commands for references (\citep{}, \citet{}, etc.), not standard LaTeX ones (\cite{}).
% Remember to include the doi and url fields in your bib database. The address field should be included for books.
% Please upload the bib file (not just the bbl one) when submitting.
 
\bibliographystyle{jasss}
\bibliography{polarization.bib} % Please set the right name for your bib file

\begin{thebibliography}{69}
\providecommand{\natexlab}[1]{#1}
\providecommand{\url}[1]{\texttt{#1}}
\providecommand{\urlprefix}{URL }
\expandafter\ifx\csname urlstyle\endcsname\relax
  \providecommand{\doi}[1]{doi:\discretionary{}{}{}#1}\else
  \providecommand{\doi}{doi:\discretionary{}{}{}\begingroup
  \urlstyle{rm}\Url}\fi

\bibitem[{Abelson(1964)}]{ABE:64}
Abelson, R.~P. (1964).
\newblock Mathematical models of the distribution of attitudes under
  controversy.
\newblock In N.~Frederiksen \& H.~Gulliksen (Eds.), \textit{Contributions To
  Mathematical Psychology}, (pp. 142--160). New York: Rinehart Winston

\bibitem[{Adamic \& Glance(2005)}]{ADA:GLA:05}
Adamic, L.~A. \& Glance, N. (2005).
\newblock The political blogosphere and the 2004 us election: divided they
  blog.
\newblock In \textit{Proceedings of the 3rd international workshop on Link
  discovery, LinkKDD '05}, (pp. 36--43)

\bibitem[{Allen(1965)}]{ALL:65}
Allen, V.~L. (1965).
\newblock Situational factors in conformity.
\newblock In L.~Berkowitz (Ed.), \textit{Advances in experimental social
  psychology}, vol.~2, (pp. 133--175). New York: Academic Press

\bibitem[{Allport(1934)}]{ALL:34}
Allport, F.~H. (1934).
\newblock The j-curve hypothesis of conforming behavior.
\newblock \textit{Journal of Social Psychology 5}, (pp. 141--183)

\bibitem[{Asch(1955)}]{ASC:55}
Asch, S.~E. (1955).
\newblock Opinions and social pressure.
\newblock \textit{Scientific American}, \textit{193}, 31--35

\bibitem[{Axelrod(1997)}]{AXE:97}
Axelrod, R. (1997).
\newblock The dissemination of culture. a model with local convergence and
  global polarization.
\newblock \textit{Journal of Conflict Resolution}, \textit{41}, 203--226.
\newblock \doi{10.1177/0022002797041002001}
\newline\urlprefix\url{http://dx.doi.org/10.1177/0022002797041002001}

\bibitem[{Brechwald \& Prinstein(2011)}]{BRE:PRI:11}
Brechwald, W.~A. \& Prinstein, M.~J. (2011).
\newblock Beyond homophily: A decade of advances in understanding peer
  influence processes.
\newblock \textit{Journal of Research on Adolescence}, \textit{21}, 166--179.
\newblock \doi{10.1111/j.1532-7795.2010.00721.x}
\newline\urlprefix\url{http://dx.doi.org/10.1111/j.1532-7795.2010.00721.x}

\bibitem[{Calais et~al.(2011)Calais, Veloso, Jr. \&
  Almeida}]{CAL:VEL:MEI:ALM:2011}
Calais, P.~H., Veloso, A., Jr., W.~M. \& Almeida, V. (2011).
\newblock From bias to opinion: a transfer-learning approach to real-time
  sentiment analysis.
\newblock In \textit{Proc. of the 17th ACM SIGKDD Conference on Knowledge
  Discovery and Data Mining}, (pp. 150--158).
\newblock \doi{10.1145/2020408.2020438}
\newline\urlprefix\url{http://dx.doi.org/10.1145/2020408.2020438}

\bibitem[{Castellano et~al.(2009{\natexlab{a}})Castellano, Fortunato \&
  Loreto}]{CAS:FOR:LOR:09}
Castellano, C., Fortunato, S. \& Loreto, V. (2009{\natexlab{a}}).
\newblock Statistical physics of social dynamics.
\newblock \textit{Reviews of Modern Physics}, \textit{81}, 591--646.
\newblock \doi{10.1103/RevModPhys.81.591}
\newline\urlprefix\url{http://dx.doi.org/10.1103/RevModPhys.81.591}

\bibitem[{Castellano et~al.(2009{\natexlab{b}})Castellano, Muñoz \&
  Pastor-Satorras}]{CAS:MUN:PAS:09}
Castellano, C., Muñoz, M.~A. \& Pastor-Satorras, R. (2009{\natexlab{b}}).
\newblock Nonlinear $q$-voter model.
\newblock \textit{Physical Review E}, \textit{80}, 041129.
\newblock \doi{10.1103/PhysRevE.80.041129}
\newline\urlprefix\url{http://dx.doi.org/10.1103/PhysRevE.80.041129}

\bibitem[{Cialdini \& Goldstein(2004)}]{CIA:GOL:04}
Cialdini, R.~B. \& Goldstein, N.~J. (2004).
\newblock Social influence: compliance and conformity.
\newblock \textit{Annual Review of Psychology}, \textit{55}, 591--621.
\newblock \doi{10.1146/annurev.psych.55.090902.142015}
\newline\urlprefix\url{http://dx.doi.org/10.1146/annurev.psych.55.090902.142015}

\bibitem[{Clifford \& Sudbury(1973)}]{CLI:SUD:73}
Clifford, P. \& Sudbury, A. (1973).
\newblock A model for spatial conflict.
\newblock \textit{Biometrika}, \textit{60}, 581--588.
\newblock \doi{10.1093/biomet/60.3.581}
\newline\urlprefix\url{http://dx.doi.org/10.1093/biomet/60.3.581}

\bibitem[{Conover et~al.(2011)Conover, Ratkiewicz, Francisco, Gonçalves,
  Flammini \& Menczer}]{CON:RAT:FRA:GON:FLA:MEN:11}
Conover, M., Ratkiewicz, J., Francisco, M., Gonçalves, B., Flammini, A. \&
  Menczer, F. (2011).
\newblock Political polarization on twitter.
\newblock In \textit{Proceedings of the Fifth International AAAI Conference on
  Weblogs and Social Media (ICWSM)}, (pp. 89--96)

\bibitem[{Crutchfield(1962)}]{CRU:62}
Crutchfield, R.~S. (1962).
\newblock Conformity and creative thinking.
\newblock In H.~E. Gruber, G.~Terrel \& M.~Wertheimer (Eds.),
  \textit{Contemporary approaches to creative thinking.}, (pp. 120--140). New
  York: Atherton

\bibitem[{DiMaggio et~al.(1996)DiMaggio, Evans \& Bryson}]{DIM:EVA:BRY:96}
DiMaggio, P., Evans, J. \& Bryson, B. (1996).
\newblock Have american's social attitudes become more polarized?
\newblock \textit{American Journal of Sociology}, \textit{102}, 690--755
\newline\urlprefix\url{http://www.jstor.org/stable/2782461}

\bibitem[{Dixit \& Weibull(2007)}]{DIX:WEI:07}
Dixit, A.~K. \& Weibull, J.~W. (2007).
\newblock Political polarization.
\newblock \textit{Proceedings of the National Academy of Sciences},
  \textit{104}, 7351--7356.
\newblock \doi{10.1073/pnas.0702071104}
\newline\urlprefix\url{http://dx.doi.org/10.1073/pnas.0702071104}

\bibitem[{French(1956)}]{FRE:56}
French, J. R.~P. (1956).
\newblock A formal theory of social power.
\newblock \textit{Psychological Review}, \textit{68}, 181--194.
\newblock \doi{10.1037/h0046123}
\newline\urlprefix\url{http://dx.doi.org/10.1037/h0046123}

\bibitem[{Galam(1997)}]{GAL:97}
Galam, S. (1997).
\newblock Rational group decision making: A random field ising model at t= 0.
\newblock \textit{Physica A}, \textit{238}, 66--80.
\newblock \doi{10.1016/S0378-4371(96)00456-6}
\newline\urlprefix\url{http://dx.doi.org/10.1016/S0378-4371(96)00456-6}

\bibitem[{Galam(2002)}]{GAL:02}
Galam, S. (2002).
\newblock Minority opinion spreading in random geometry.
\newblock \textit{European Physical Journal B}, \textit{25}, 403--406.
\newblock \doi{10.1140/epjb/e20020045}
\newline\urlprefix\url{http://dx.doi.org/10.1140/epjb/e20020045}

\bibitem[{Galam(2004)}]{GAL:04}
Galam, S. (2004).
\newblock Contrarian deterministic effects on opinion dynamics:``the hung
  elections scenario''.
\newblock \textit{Physica A}, \textit{333}, 453--460.
\newblock \doi{10.1016/j.physa.2003.10.041}
\newline\urlprefix\url{http://dx.doi.org/10.1016/j.physa.2003.10.041}

\bibitem[{Galam \& Jacobs(2007)}]{GAL:JAC:07}
Galam, S. \& Jacobs, F. (2007).
\newblock The role of inflexible minorities in the breaking of democratic
  opinion dynamics.
\newblock \textit{Physica A}, \textit{381}, 366--376.
\newblock \doi{10.1016/j.physa.2007.03.034}
\newline\urlprefix\url{http://dx.doi.org/10.1016/j.physa.2007.03.034}

\bibitem[{Galam \& Moscovici(1991)}]{GAL:MOS:91}
Galam, S. \& Moscovici, S. (1991).
\newblock Towards a theory of collective phenomena: Consensus and attitude
  changes in groups.
\newblock \textit{European Journal of Social Psychology}, \textit{21}, 49--74.
\newblock \doi{10.1002/ejsp.2420210105}
\newline\urlprefix\url{http://dx.doi.org/10.1002/ejsp.2420210105}

\bibitem[{Griskevicius et~al.(2006)Griskevicius, Goldstein, Mortensen, Cialdini
  \& Kenrick}]{GRI:GOL:MOR:CIA:KEN:06}
Griskevicius, V., Goldstein, N.~J., Mortensen, C.~R., Cialdini, R.~B. \&
  Kenrick, D.~T. (2006).
\newblock Going along versus going alone: When fundamental motives facilitate
  strategic (non)conformity.
\newblock \textit{Journal of Personality and Social Psychology}, \textit{91},
  281--294.
\newblock \doi{10.1037/0022-3514.91.2.281}
\newline\urlprefix\url{http://dx.doi.org/10.1037/0022-3514.91.2.281}

\bibitem[{Gruzd \& Roy(2014)}]{GRU:ROY:14}
Gruzd, A. \& Roy, J. (2014).
\newblock Investigating political polarization on twitter: A canadian
  perspective.
\newblock \textit{Policy and Internet}, \textit{6}, 28--45.
\newblock \doi{10.1002/1944-2866.POI354}
\newline\urlprefix\url{http://dx.doi.org/10.1002/1944-2866.POI354}

\bibitem[{Guerra et~al.(2013)Guerra, Jr., Cardie \&
  Kleinberg}]{GUE:MEI:CAR:KLE:13}
Guerra, P. H.~C., Jr., W.~M., Cardie, C. \& Kleinberg, R. (2013).
\newblock A measure of polarization on social media networks based on community
  boundaries.
\newblock In \textit{Proceedings of the Seventh International AAAI Conference
  on Web and Social Media}, (pp. 215--224)
\newline\urlprefix\url{http://www.aaai.org/ocs/index.php/ICWSM/ICWSM13/paper/view/6104}

\bibitem[{Harary(1959)}]{HAR:59}
Harary, F. (1959).
\newblock A criterion for unanimity in french's theory of social power.
\newblock In D.~Cartwright (Ed.), \textit{Studies in Social Power}, (pp.
  168--182). Ann Arbor: Institute for Social Research

\bibitem[{Hengselmann \& Krause(2002)}]{HEN:KRA:02}
Hengselmann, R. \& Krause, U. (2002).
\newblock Opinion dynamics and bounded confidence: models, analysis and
  simulation.
\newblock \textit{Journal of Artificial Societies and Social Simulation},
  \textit{5}(3)
\newline\urlprefix\url{http://jasss.soc.surrey.ac.uk/5/3/2.html}

\bibitem[{Holley \& Liggett(1975)}]{HOL:LIG:75}
Holley, R. \& Liggett, T. (1975).
\newblock Ergodic theorems for weakly interacting systems and the voter model.
\newblock \textit{Annals of Probability}, \textit{3}, 643--663

\bibitem[{Huckfeldt(2007)}]{HUC:07}
Huckfeldt, R. (2007).
\newblock Unanimity, discord, and the communication of public opinion.
\newblock \textit{American Journal of Political Science}, \textit{51}, 978--995

\bibitem[{Isenberg(1986)}]{ISE:86}
Isenberg, D.~J. (1986).
\newblock Group polarization: A critical review and meta-analysis.
\newblock \textit{Journal of Personality and Social Psychology}, \textit{50},
  1141--1151.
\newblock \doi{10.1037/0022-3514.50.6.1141}
\newline\urlprefix\url{http://dx.doi.org/10.1037/0022-3514.50.6.1141}

\bibitem[{Jędrzejewski et~al.(2016)Jędrzejewski, Sznajd-Weron \&
  Szwabiński}]{JED:SZN:SZW:16}
Jędrzejewski, A., Sznajd-Weron, K. \& Szwabiński, J. (2016).
\newblock Mapping the q-voter model: From a single chain to complex networks.
\newblock \textit{Physica A}, \textit{446}, 110--119.
\newblock \doi{10.1016/j.physa.2015.11.005}
\newline\urlprefix\url{http://dx.doi.org/10.1016/j.physa.2015.11.005}

\bibitem[{Klemm et~al.(2003)Klemm, Eguíluz, Toral \&
  Miguel}]{KLE:EGU:TOR:MIG:03}
Klemm, K., Eguíluz, V.~M., Toral, R. \& Miguel, M.~S. (2003).
\newblock Nonequilibrium transitions in complex networks: A model of social
  interaction.
\newblock \textit{Physical Review E}, \textit{67}, 026120.
\newblock \doi{10.1103/PhysRevE.67.026120}
\newline\urlprefix\url{http://dx.doi.org/10.1103/PhysRevE.67.026120}

\bibitem[{Krech et~al.(1962)Krech, Crutchfield \& Ballachey}]{KRE:CRU:BAL:62}
Krech, D., Crutchfield, R.~S. \& Ballachey, E.~L. (1962).
\newblock \textit{Individual in society: A textbook of social psychology.}
\newblock New York: McGraw-Hill

\bibitem[{Krizan \& Baron(2007)}]{KRI:BAR:07}
Krizan, Z. \& Baron, R.~S. (2007).
\newblock Group polarization and choice-dilemmas: how important is
  self-categorization?
\newblock \textit{European Journal of Social Psychology}, \textit{37},
  191--201.
\newblock \doi{10.1002/ejsp.345}
\newline\urlprefix\url{http://dx.doi.org/10.1002/ejsp.345}

\bibitem[{Leifeld(2014)}]{LEI:14}
Leifeld, P. (2014).
\newblock Polarization of coalitions in an agent-based model of political
  discourse.
\newblock \textit{Computational Social Networks}, \textit{1}, 1--22.
\newblock \doi{10.1186/s40649-014-0007-y}
\newline\urlprefix\url{http://dx.doi.org/10.1186/s40649-014-0007-y}

\bibitem[{Lewenstein et~al.(1992)Lewenstein, Nowak \& Latané}]{LEW:NOW:LAT:92}
Lewenstein, M., Nowak, A. \& Latané, B. (1992).
\newblock Statistical mechanics of social impact.
\newblock \textit{Physical Review A}, \textit{45}, 763--776.
\newblock \doi{10.1103/PhysRevA.45.763}
\newline\urlprefix\url{http://dx.doi.org/10.1103/PhysRevA.45.763}

\bibitem[{Livne et~al.(2011)Livne, Simmons, Adar \&
  Adamic}]{LIV:SIM:ADA:ADA:11}
Livne, A., Simmons, M.~P., Adar, E. \& Adamic, L.~A. (2011).
\newblock The party is over here: Structure and content in the 2010 election.
\newblock In \textit{Proceedings of the 5th International AAAI Conference on
  Weblogs and Social Media}, (pp. 201--208)
\newline\urlprefix\url{https://www.aaai.org/ocs/index.php/ICWSM/ICWSM11/paper/viewFile/2852/3272}

\bibitem[{Macy et~al.(2003)Macy, Kitts, Flache \& Benard}]{MAC:KIT:FLA:BEN:03}
Macy, M.~W., Kitts, J., Flache, A. \& Benard, S. (2003).
\newblock Polarization and dynamic networks. a hopfield model of emergent
  structure.
\newblock In R.~Breiger, K.~Carley \& P.~Pattison (Eds.), \textit{Dynamic
  Social Network Modeling and Analysis: Workshop Summary and Papers.}, (pp.
  162--173). Washington DC: The National Academies Press
\newline\urlprefix\url{http://www.nap.edu/read/10735/chapter/11}

\bibitem[{Macy \& Willer(2002)}]{MAC:WIL:02}
Macy, M.~W. \& Willer, R. (2002).
\newblock From factors to actors: Computational sociology and agent-based
  modeling.
\newblock \textit{Annual Review of Sociology}, \textit{28}, 143--166
\newline\urlprefix\url{http://www.jstor.org/stable/3069238}

\bibitem[{Maoz(2006)}]{MAO:06}
Maoz, Z. (2006).
\newblock Network polarization, network interdependence, and international
  conflict.
\newblock \textit{Journal of Peace Research}, \textit{43}, 1816--2002.
\newblock \doi{10.1177/0022343306065720}
\newline\urlprefix\url{http://dx.doi.org/10.1177/0022343306065720}

\bibitem[{Marvel et~al.(2011)Marvel, Kleinberg, Kleinberg \&
  Strogatz}]{MAR:KLE:KLE:STR:11}
Marvel, S.~A., Kleinberg, J., Kleinberg, R.~D. \& Strogatz, S.~H. (2011).
\newblock Continuous-time model of structural balance.
\newblock \textit{Proceedings of the National Academy of Sciences},
  \textit{108}, 1771--1776.
\newblock \doi{10.1073/pnas.1013213108}
\newline\urlprefix\url{http://dx.doi.org/10.1073/pnas.1013213108}

\bibitem[{McCright \& Dunlap(2011)}]{MCC:DUN:11}
McCright, A.~M. \& Dunlap, R.~E. (2011).
\newblock The politization of climate change and polarization in the american
  public's views of global warming.
\newblock \textit{The Sociological Quaterly}, \textit{52}, 2001--2010.
\newblock \doi{10.1111/j.1533-8525.2011.01198.x}
\newline\urlprefix\url{http://dx.doi.org/10.1111/j.1533-8525.2011.01198.x}

\bibitem[{Mckeown \& Sheely(2006)}]{MCK:SHE:06}
Mckeown, G. \& Sheely, N. (2006).
\newblock Mass media and polarisation processes in the bounded confidence model
  of opinion dynamics.
\newblock \textit{Journal of Artificial Societies and Social Simulation},
  \textit{9}(1), 11
\newline\urlprefix\url{http://jasss.soc.surrey.ac.uk/9/1/11.html}

\bibitem[{McPherson et~al.(2001)McPherson, Smith-Lovin \&
  Cook}]{MCP:SMI:COO:01}
McPherson, M., Smith-Lovin, L. \& Cook, J.~M. (2001).
\newblock Birds of a feather: Homophily in social networks.
\newblock \textit{Annual Review of Sociology}, \textit{27}, 415--444.
\newblock \doi{10.1146/annurev.soc.27.1.415}
\newline\urlprefix\url{http://dx.doi.org/10.1146/annurev.soc.27.1.415}

\bibitem[{Mouw \& Sobel(2001)}]{MOU:SOB:01}
Mouw, T. \& Sobel, M. (2001).
\newblock Culture wars and opinion polarization: The case of abortion.
\newblock \textit{American Journal of Sociology}, \textit{106}, 913--943

\bibitem[{Myers(2013)}]{MYE:13}
Myers, D.~G. (2013).
\newblock \textit{Social Psychology.}
\newblock New York: Freeman Press, 11th edn.

\bibitem[{Mäs \& Flache(2013)}]{MAS:FLA:13}
Mäs, M. \& Flache, A. (2013).
\newblock Differentiation without distancing. explaining bi-polarization of
  opinions without negative influence.
\newblock \textit{PLoS ONE}, \textit{8}(11), e74516.
\newblock \doi{10.1371/journal.pone.0074516}
\newline\urlprefix\url{http://dx.doi.org/10.1371/journal.pone.0074516}

\bibitem[{Nail(1986)}]{NAI:86}
Nail, P.~R. (1986).
\newblock Toward an integration of some models and theories of social response.
\newblock \textit{Psychological Bulletin}, \textit{100}, 190--206.
\newblock \doi{10.1037/0033-2909.100.2.190}
\newline\urlprefix\url{http://dx.doi.org/10.1037/0033-2909.100.2.190}

\bibitem[{Nail et~al.(2013)Nail, Domenico \& MacDonald}]{NAI:DOM:MAC:13}
Nail, P.~R., Domenico, S.~I. \& MacDonald, G. (2013).
\newblock Proposal of a double diamond model of social response.
\newblock \textit{Review of General Psychology}, \textit{17}, 1--19.
\newblock \doi{10.1037/a0030997}
\newline\urlprefix\url{http://dx.doi.org/10.1037/a0030997}

\bibitem[{Nail et~al.(2000)Nail, MacDonald \& Levy}]{NAI:DON:LEV:00}
Nail, P.~R., MacDonald, G. \& Levy, D.~A. (2000).
\newblock Proposal of a four dimensional model of social response.
\newblock \textit{Psychological Bulletin}, \textit{126}, 454--470.
\newblock \doi{10.1037/0033-2909.126.3.454}
\newline\urlprefix\url{http://dx.doi.org/10.1037/0033-2909.126.3.454}

\bibitem[{Newman(2006)}]{NEW:06}
Newman, M. E.~J. (2006).
\newblock Modularity and community structure in networks.
\newblock \textit{Proceedings of the National Academy of Sciences},
  \textit{103}, 8577--8582.
\newblock \doi{10.1073/pnas.0601602103}
\newline\urlprefix\url{http://dx.doi.org/10.1073/pnas.0601602103}

\bibitem[{Nyczka \& Sznajd-Weron(2013)}]{NYC:SZN:13}
Nyczka, P. \& Sznajd-Weron, K. (2013).
\newblock Anticonformity or independence?—insights from statistical physics.
\newblock \textit{Journal of Statistical Physics}, \textit{151}, 174--202.
\newblock \doi{10.1007/s10955-013-0701-4}
\newline\urlprefix\url{http://dx.doi.org/10.1007/s10955-013-0701-4}

\bibitem[{Nyczka et~al.(2012)Nyczka, Sznajd-Weron \& Cislo}]{NYC:SZN:CIS:12}
Nyczka, P., Sznajd-Weron, K. \& Cislo, J. (2012).
\newblock Phase transitions in the q-voter model with two types of stochastic
  driving.
\newblock \textit{Physical Review E}, \textit{86}, 011105.
\newblock \doi{10.1103/PhysRevE.86.011105}
\newline\urlprefix\url{http://dx.doi.org/10.1103/PhysRevE.86.011105}

\bibitem[{Przybyła et~al.(2014)Przybyła, Sznajd-Weron \&
  Weron}]{PRZ:SZN:WER:14}
Przybyła, P., Sznajd-Weron, K. \& Weron, R. (2014).
\newblock Diffusion of innovation within an agent-based model: Spinsons,
  independence and advertising.
\newblock \textit{Advances in Complex Systems}, \textit{17}, 1450004.
\newblock \doi{10.1142/S0219525914500040}
\newline\urlprefix\url{http://dx.doi.org/10.1142/S0219525914500040}

\bibitem[{Salzarulo(2006)}]{SAL:06}
Salzarulo, L. (2006).
\newblock A continuous opinion dynamics model based on the principle of
  meta-contrast.
\newblock \textit{Journal of Artificial Societies and Social Simulation},
  \textit{9}(1), 13
\newline\urlprefix\url{http://jasss.soc.surrey.ac.uk/9/1/13.html}

\bibitem[{Sherif(1935)}]{SHE:35}
Sherif, M. (1935).
\newblock A study of some social factors in perception.
\newblock \textit{Archives of Psychology}, \textit{27}, 1--60

\bibitem[{Sood et~al.(2008)Sood, Antal \& Redner}]{SOO:ANT:RED:08}
Sood, V., Antal, T. \& Redner, S. (2008).
\newblock Voter models on heterogeneous networks.
\newblock \textit{Physical Review E}, \textit{77}, 041121.
\newblock \doi{10.1103/PhysRevE.77.041121}
\newline\urlprefix\url{http://dx.doi.org/10.1103/PhysRevE.77.041121}

\bibitem[{Stouffer et~al.(1950)Stouffer, Guttman, Suchman, Lazarsfeld, Star \&
  Clausen}]{STO:GUT:SUC:LAZ:STA:CLA:50}
Stouffer, S., Guttman, L., Suchman, E.~A., Lazarsfeld, P., Star, S. \& Clausen,
  J. (Eds.) (1950).
\newblock \textit{Studies in Social Psychology in World War II}, vol.~4.
\newblock Princeton: Princeton University Press

\bibitem[{Sunstein(2002)}]{SUN:02}
Sunstein, C.~R. (2002).
\newblock The law of group polarization.
\newblock \textit{Journal of Political Philosophy}, \textit{10}, 175--195.
\newblock \doi{10.1111/1467-9760.00148}
\newline\urlprefix\url{http://dx.doi.org/10.1111/1467-9760.00148}

\bibitem[{Sznajd-Weron \& Sznajd(2000)}]{SZN:SZN:00}
Sznajd-Weron, K. \& Sznajd, J. (2000).
\newblock Opinion evolution in closed community.
\newblock \textit{International Journal of Modern Physics C}, \textit{11},
  1157.
\newblock \doi{10.1142/S0129183100000936}
\newline\urlprefix\url{http://dx.doi.org/10.1142/S0129183100000936}

\bibitem[{Sznajd-Weron et~al.(2014{\natexlab{a}})Sznajd-Weron, Szwabiński \&
  Weron}]{SZN:SZW:WER:14}
Sznajd-Weron, K., Szwabiński, J. \& Weron, R. (2014{\natexlab{a}}).
\newblock Is the person-situation debate important for agent-based modeling and
  vice-versa?
\newblock \textit{PLoS ONE}, \textit{9}(11), e112203.
\newblock \doi{10.1371/journal.pone.0112203}
\newline\urlprefix\url{http://dx.doi.org/10.1371/journal.pone.0112203}

\bibitem[{Sznajd-Weron et~al.(2014{\natexlab{b}})Sznajd-Weron, Szwabiński,
  Weron \& Weron}]{SZN:SZW:WER:WER:14}
Sznajd-Weron, K., Szwabiński, J., Weron, R. \& Weron, T. (2014{\natexlab{b}}).
\newblock Rewiring the network. what helps an innovation to diffuse?
\newblock \textit{Journal of Statistical Mechanics}, \textit{2014}, P03007.
\newblock \doi{10.1088/1742-5468/2014/03/P03007}
\newline\urlprefix\url{http://dx.doi.org/10.1088/1742-5468/2014/03/P03007}

\bibitem[{Sznajd-Weron et~al.(2011)Sznajd-Weron, Tabiszewski \&
  Timpanaro}]{SZN:TAB:TIM:11}
Sznajd-Weron, K., Tabiszewski, M. \& Timpanaro, A. (2011).
\newblock Phase transition in the sznajd model with independence.
\newblock \textit{Europhysics Letters}, \textit{96}, 48002.
\newblock \doi{10.1209/0295-5075/96/48002}
\newline\urlprefix\url{http://dx.doi.org/10.1209/0295-5075/96/48002}

\bibitem[{Traag et~al.(2013)Traag, Dooren \& Leenheer}]{TRA:DOO:LEE:13}
Traag, V.~A., Dooren, P.~V. \& Leenheer, P.~D. (2013).
\newblock Dynamical models explaining social balance and evolution of
  cooperation.
\newblock \textit{PLoS ONE}, \textit{8}(4), e60063.
\newblock \doi{10.1371/journal.pone.0060063}
\newline\urlprefix\url{http://dx.doi.org/10.1371/journal.pone.0060063}

\bibitem[{Walton(1991)}]{WAL:91}
Walton, D. (1991).
\newblock Bias, critical doubt, and fallacies.
\newblock \textit{Argumentation and Advocacy}, \textit{28}, 1--22

\bibitem[{Watts \& Dodds(2007)}]{WAT:DOD:07}
Watts, D.~J. \& Dodds, P.~S. (2007).
\newblock Influentials, networks, and public opinion formation.
\newblock \textit{Journal of Consumer Research}, \textit{34}, 441--458.
\newblock \doi{10.1086/518527}
\newline\urlprefix\url{http://dx.doi.org/10.1086/518527}

\bibitem[{Waugh et~al.(2011)Waugh, Pei, Fowler, Mucha \&
  Porter}]{WAU:PEI:FOW:MUC:POR:11}
Waugh, A., Pei, L., Fowler, J.~H., Mucha, P.~J. \& Porter, M.~A. (2011).
\newblock Party polarization in congress: a network science approach.
\newline\urlprefix\url{http://arxiv.org/pdf/0907.3509v3.pdf}

\bibitem[{Willis(1963)}]{WIL:63}
Willis, R.~H. (1963).
\newblock Two dimensions of conformity-nonconformity.
\newblock \textit{Sociometry}, \textit{1963}, 499--513

\bibitem[{Zachary(1977)}]{ZAC:77}
Zachary, W. (1977).
\newblock An information flow model for conflict and fission in small groups.
\newblock \textit{Journal of Anthropological Research}, \textit{33}, 452--473
\newline\urlprefix\url{http://www.jstor.org/stable/3629752}

\end{thebibliography}

%%%%%%%%%%%%%%%%%%%%%%%%%%%%%%%%%%%%%%%%%%%%%%

\end{document}